\documentclass[useAMS,usenatbib]{mnras}
\usepackage{graphicx} 
\usepackage{pslatex}
\usepackage{bm}
\usepackage[table]{xcolor} 
\usepackage{soul} 

\title[Fast Radio Bursts and cosmological tests]
{Fast Radio Bursts and cosmological tests}

\author[M. Jaroszynski]
{M. Jaroszynski$^{1}$\thanks{E-mail:mj@astrouw.edu.pl}\\
$^{1}$University of Warsaw Observatory, Al. Ujazdowskie 4, 00-478 Warsaw,
Poland}

\begin{document}

\date{Accepted; Received ;}

\pagerange{\pageref{1}--\pageref{9}} \pubyear{2018}

\definecolor{HighlightColor}{HTML}{00FFCF}
\sethlcolor{HighlightColor}
\newcommand{\modi}[1]{\hl{#1}}

\maketitle

\label{firstpage}

\begin{abstract}
We consider future cosmological tests based on observations of Fast
Radio Bursts (FRBs). We use Illustris Simulation to realistically
estimate the scatter in the dispersion measure ($DM$) of FRBs caused by
the inhomogeneous distribution of ionized gas in {\rm the Intergalactic
Medium (IGM)}. We find $\sim$ {\rm 13}\%  scatter in $DM$ to a source 
at $z=1$ and $\sim$ {\rm 7}\% at $z=3$ (one sigma). The
distribution of $DM$ is {\rm close} to Gaussian.
We simulate samples of FRBs and examine their applicability to simple
cosmological tests. Our calculations show that using a sample of 100 FRBs
and fixing cosmological model one can find the redshift and sample averaged
fraction of ionized gas with $\sim$ {\rm 1}\% uncertainty. Finding the ionized
fraction with  $\sim$ 1\% accuracy at few different epochs would require
$\sim 10^4$ FRBs with known redshifts. Because $DM$ is proportional to the
product of ionized fraction, baryon density and the Hubble constant it
is impossible to constrain these parameters separately with FRBs.
Constraints on cosmological densities are possible in a flat
$\Lambda$CDM model but give uninterestingly low accuracy. Using FRBs
with other type of data improves the constraints, but the role of FRBs
is not crucial. Thus constraints on the distribution of ionized gas are
probably the most promising application of FRBs which allow for
"tomography" if sources redshifts are known, as opposed to measuring
$\tau_e$ or $y$ parameters with CMB observations.

\end{abstract}

\begin{keywords}
radio continuum: transients -- cosmological parameters
\end{keywords}

\section{Introduction}

The discovery of fast radio bursts (FRBs) \citep{lor07}, large dispersion
measures ($DM$) of some of them \citep{ten17}, and measured redshifts 
\citep{chat17, ten17} make them a class of phenomena which (at least in
part) may be happening at
cosmological distances. Their application to cosmological tests has been
proposed by several authors \citep{zhou14, gao14, lor16, yw17}, to cite few.

In a recent article \citep{rav17} the properties of known FRBs are
discussed. They seem to be a nonuniform class of objects with
complicated characteristics and the extragalactic origin of $DM$ is likely
only in a part of the sources.  The number of known FRBs is far too low to
study their distribution on the sky or make a meaningful $\lg F$ -- $\lg N$
test, which could disprove the cosmological hypothesis (compare
\citealt{BP} for the case of gamma ray bursts and \citealt{katz17} for FRBs).
We assume, that there is a cosmological population of FRBs and that the improved
instruments (e.g. Square Kilometer Array) will allow to study a large number of
these events.

Locally the observed $DM$ is proportional to the column density of free electrons
along the line of sight (LOS) to the source. It is precisely measured for
pulsars (see e.g. \citealt{bil16}). It is used to find distances to the
sources independently of any {\it cosmic ladder}. 
$DM$ measurements are based on the fact that lower frequency radio signals
travel with lower velocities in plasma, so they are delayed  
as compared to a higher frequency signals. 
The  delay can be measured. In an expanding universe the contributions
from free electron column density at small redshift interval $z$ --
$z+\Delta z$ scale with the redshift factor as $\propto 1/(1+z)$. On the
other hand one can expect the density of free electrons $n_e\propto (1+z)^3$
which makes the
contributions from high redshift dominating. Thus there is no direct
proportionality between $DM$ and the proper distance to the source 
even in an uniform universe.

\cite{sh18} model the distribution of free electrons in space based
on the observations of absorption lines in QSO spectra and the relations
between the neutral and ionized fractions of the gas at different redshifts.
They also estimate the redshifts of the known FRBs using a uniform flat
universe model, showing that in five of twenty five considered cases
$z>1$. 

We are simulating cosmological tests based on possible future samples of
FRBs with known dispersion measures and redshifts.
In our approach we calculate $DM$ along many LOS to distant sources 
employing the results of the Illustris Simulation 
\citep{vog14a, vog14b} which are publicly available \citep{nel15} and
give the spatial distribution of ionized gas for a discrete set of
redshifts. 
{\rm Cosmological simulations were used in the past by \cite{mcq14} 
 to estimate
the variance of the dispersion measure  due to the inhomogeneity of
the gas distribution in space and by \cite{dol15}, who considered various
aspects of FRBs including the large scale structure influence on the
derived properties of sources population.}

In the next section we describe our method to calculate $DM$ to a source
at given redshift using Illustris data. 
In Sec.~3 we describe a possible cosmological test based on a simulated 
sample of a hundred FRBs with measured hosts redshifts, {\rm
concentrating on possible estimates of the fraction of ionized gas in space.}
In Sec.~4 we use the SN Ia {\it Union} sample 
\citealt{kow08} to {\rm investigate the role of other data on the
applicability of FRBs to cosmological tests.}
Discussion follows in the last section.

\section{Calculation of the dispersion measure in a nonuniform universe
model}

\subsection{The method}

We are using the results of Illustris Simulation to 
describe the evolving ionized gas distribution in space. 
The Simulation provides the history of
the structure formation in a cube of 75 $\mathrm{Mpc}/h$ edge length,
including the distribution of free electrons. To save the storage space and
the calculations time we use Illustris-3, which has the lowest spatial
resolution. 

The dispersion measure for a source at the
redshift $z_\mathrm{S}$, neglecting the contribution from our Galaxy, the
source itself , and its host galaxy, {\rm can be calculated as an integral
along the path from the source to the observer (compare
\citealt{dz14}, \citealt{zhou14}, \citealt{gao14}):}
\begin{equation}
DM(z_S)=\int_0^{z_S}~\frac{n_e(z)}{1+z}\frac{dl_\mathrm{prop}}{dz}dz
=\int_0^{z_S}~\frac{c/H_0~n_e(z)dz}{(1+z)^2h(z)}\\
\label{defDM1}
\end{equation}
where $n_e$ is the concentration of free electrons along the LOS, 
$dl_\mathrm{prop}$ is the proper distance differential, 
$z$ - the redshift, $c$ - the speed of light, 
and $H_0=100~h~\mathrm{km/s/Mpc}$ - the Hubble constant.
In a $\Lambda$CDM model: 
\begin{equation}
h(z)=\sqrt{\Omega_M(1+z)^3+\Omega_K(1+z)^2+\Omega_\Lambda}
\label{defDM2}
\end{equation}
where $\Omega_M$ is the matter density parameter (baryons and dark matter),
$\Omega_\Lambda$ - dark energy density parameter, 
$\Omega_K\equiv 1-\Omega_M-\Omega_\Lambda$ - the curvature parameter, 
and $h(z)\equiv H(z)/H_0$ gives the rate of
expansion as a function of $z$. {\rm  $DM$ here represents its inter-galaxy
medium part ($DM_{IGM}$) but we omit this subscript for compactness.
In Illustris the cosmological  parameters have the following values
(denoted by an extra "I" in the subscripts):
$H_{0I}=70.4$ km s$^{-1}$ Mpc$^{-1}$, $\Omega_{MI}=0.2726$, 
$\Omega_{\Lambda I}=0.7274$, and the baryon density parameter 
$\Omega_{BI}=0.0456$.}

{rm  The characteristic present number density of electrons in a uniform
universe model with hydrogen and helium mass fractions $X$, $Y$,
assuming complete ionization is given as:
\begin{equation}
n_{e0}=\left(X+\frac{1}{2}Y\right)
\Omega_B\frac{3H_0^2}{8\pi G}\frac{1}{m_H}
=2.22\times 10^{-7}~\mathrm{cm}^{-3}
\label{nel0}
\end{equation}
The numeric value is calculated for the Illustris model parameters with
the primordial abundances  $X=0.75$ and $Y=0.25$.}

{\rm The averaged free electron concentration {rm in the IGM} 
$\left<n_e\right>(z)$ depends on {rm the fraction of gas mass belonging to
the IGM $f_{IGM}$,}
on the the gas chemical composition, and the ionization state
of all species. 
{rm We use a redshift dependent factor $f_e(z)\equiv
f_{IGM}(z)f_{ion}(z)$, where $f_{ion}$ describes the effects of 
chemical composition and ionization state of the gas. For a primordial
chemical composition and complete ionization $f_{ion}=1$ by definition.}
In general the averaged
electron concentration {\rm in the IGM }can be expressed using $f_e(z)$ as: }
\begin{equation}
\left<n_e\right>(z)\equiv f_e(z)n_{e0}(1+z)^3
\label{deffe}
\end{equation}

{\rm It is convenient to rewrite Eq.~\ref{defDM1} in the form, which
separates the terms related to the concentration fluctuations from terms
depending on the uniform cosmological model:}
\begin{equation}
DM(z_S)=\int_0^{z_S}~n_{e0}f_e(z)(1+\delta_3)
\frac{c/H_0~(1+z)dz}{h(z)}\\
\label{defDM3}
\end{equation}
{\rm where $\delta_3$ is the 3D electron density fluctuation 
($n_e=\left<n_e\right>(z)(1+\delta_3)$) - the only variable under the
integral which depends on the spatial coordinates.}

To employ the results of Illustris we imagine that our line of sight
goes through several simulation cubes corresponding to different epochs
(redshift ranges). The matter distribution is correlated on scales of
tens of megaparsecs \citep{gel87} and modeling the line of sight as a sum of
sections, each belonging to a single cube takes these correlations into
account. On the other hand there should be no strong correlations on larger
scales and choosing at random photon paths through each cube
guarantees that. {\rm For simplicity we consider photon paths parallel to one
of the simulation cube edges, which} saves programming and calculation time.
(Compare \citealp{carbone08} 
about shifting and rotating simulation cubes at different epochs to get
better approximation of statistical uniformity and isotropy of the
simulated matter distribution.)

{\rm We construct a sequence of simulation cubes on the photon path using
the condition:
\begin{equation}
\int_{z_{i-1}}^{z_i}~(1+z)\frac{dl_{prop}}{dz}~dz=
\int_{z_{i-1}}^{z_i}~\frac{c/H_0~dz}{h(z)} = 75~\mathrm{Mpc}/h
\label{sequence}
\end{equation}
where the integral gives the comoving distance along the path. We start
from $z_0=0$ and calculate the other $z_i$ recursively. The redshift
intervals corresponding to the photon travel through a single cube are small
($z_i-z_{i-1}\ll 1+z_i$) so we can neglect the effects of cosmic
expansion/evolution and use distribution of electrons given by $\delta_3$
corresponding to the averaged redshift $z_{av,i}\equiv (z_{i-1}+z_i)/2$. 
Projecting the 3D distribution of the electron density fluctuations
$\delta_3(z_{av,i})$ along one of the simulation cube edges we get the 2D
distribution of the fluctuations in surface electron density 
$\delta_2(z_{av,i})$. The dispersion measure to a point at the 
redshift $z_i$ can now be expressed as:
\begin{equation}
DM(z_i)=\sum_{j=1}^i~(1+\delta_2(z_{av,j}))
\int_{z_{j-1}}^{z_j}~n_{e0}f_e(z)\frac{(1+z)c/H_0dz}{h(z)}
\label{deltaDM}
\end{equation}
For other source redshifts we get $DM(z_S)$ by interpolation. The
averaged dispersion measure $\left<DM(z_S)\right>$ can be calculated by
substituting $\delta_2(z_{av,j})=0$ into Eq.~\ref{deltaDM}.
}

\begin{figure}
\includegraphics[width=\columnwidth]{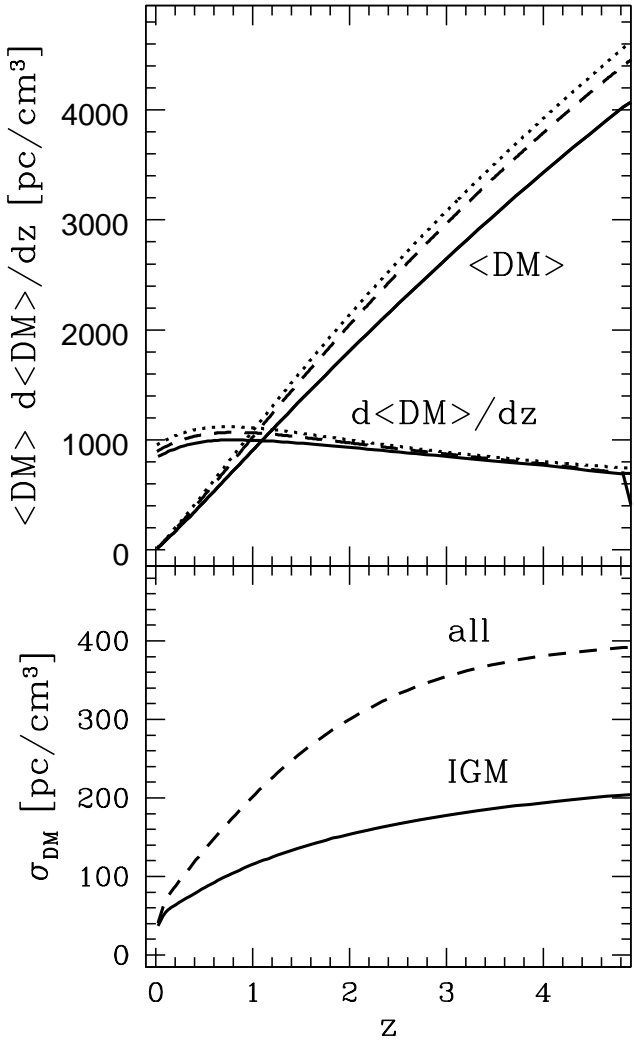}
\caption{(Upper panel) The dependence of the averaged dispersion measure on the
redshift and its derivative.
Solid lines show the relations corresponding to the averaged
electron density {\rm in the IGM alone} taken from the Illustris simulation,
{\rm the dashed lines take into account all free electrons according 
to Illustris}, 
and the dotted lines are calculated under the assumption that all 
baryonic matter has the form of completely ionized gas with primordial 
composition. 
(Lower panel) The standard deviation of the dispersion measure is shown.
Drawing conventions follow the upper panel
}
\label{avgDM}
\end{figure}

{\rm In Fig.~\ref{avgDM} we show the redshift dependence of the dispersion
measure in a model with uniform electron density and
cosmological parameters corresponding to the Illustris simulation.
The solid line shows the
relation for averaged electron density {\rm in the IGM} 
taken from Illustris (see the
next subsection). {\rm The dashed line shows the dependence in a case,
when all free electrons, regardless of their position relative to
dark matter haloes, are included.} 
Calculation with the same cosmological parameters but
assuming $f_e(z)=1$ gives the dotted line. The scatter 
in the dispersion measure $\sigma_{DM}$ resulting from the nonuniform 
electron distribution is shown in the lower panel. For $z=1$ we get 
$\sigma_{DM}=115~\mathrm{pc}~\mathrm{cm}^{-3}$ {\rm about two times
smaller than} the
result of \cite{mcq14} (his Fig.~1, lower panel) based on simulations. 
{\rm On the other hand our result based on the distribution of all
electrons is in agreement with his value. }

\begin{figure}
\includegraphics[width=\columnwidth]{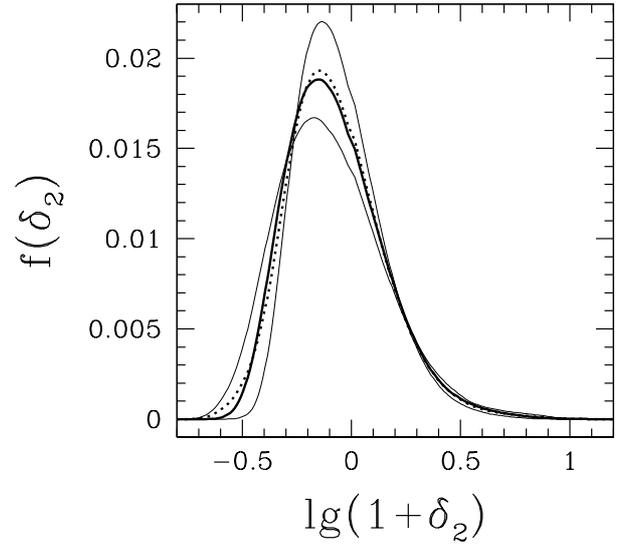}
\caption{The probability distributions $f_{0.5}(\delta_2)$ (thick 
solid line), $f_0(\delta_2)$, $f_1(\delta_2)$ (thin solid lines), 
and interpolated  distribution $f_{01}(\delta_2)$ (dotted) plotted as 
functions of $\lg(1+\delta_2)$. See the text for details.
}
\label{interpolated}
\end{figure}

\subsection{Distribution of ionized gas from Illustris simulation}

We have downloaded the 3D maps of the distribution of matter in 
the Illustris-3 simulation cube corresponding to the epochs 
$z=0$, {\rm 0.5}, 1, 2, 3, 4, 5, 5.5, and 6. 
The simulation gives the positions, masses, densities,
and ionization states of $\approx 455^3$ gas cells present at the
beginning of calculations. Some of them become stars/wind particles
during the evolution. The distribution of 
dark matter is given similarly as positions and masses of another 
$455^3$ dark matter particles.

{\rm \cite{hai16} analyze large-scale mass distribution in the Illustris
simulation, dividing the space into regions of voids, filaments, and
haloes based on the local dark matter density. A region with 
$\rho_{DM}\ge 15\rho_c$ (dark matter density fifteen times higher than
critical) is treated as belonging to a halo. 
The haloes occupy a tiny fraction of the volume (0.16\% today), but 49\%
of the dark matter and 23\% of the baryons can be found there.
}

\begin{table}

  \caption{Distribution of $DM$ [$\mathrm{pc}~\mathrm{cm}^{-3}$]
      for $z=1$, 2, 3, 4, and 5}
  \begin{tabular}{crcccc}
  \hline
  $z$ & $\left<DM\right>$ & $\sigma_{DM}$ & $\gamma_3$ & $\gamma_4$\\
\hline
 1 &  905. & 115. & 0.64 & 3.79 \\
 2 & 1803. & 154. & 0.48 & 3.44 \\
 3 & 2649. & 177. & 0.39 & 3.30 \\
 4 & 3431. & 194. & 0.34 & 3.23 \\
 5 & 4136. & 204. & 0.31 & 3.19 \\
 \hline
\end{tabular}
\label{cumulDM}
\end{table}

{\rm The 3D electron density fluctuations $\delta_3$ are defined by the
distribution of gas elements. Each gas cell has defined mass, mean
density, and electron abundance, so their characteristic size and
electron number can be calculated. {\rm The total mass density $\rho$ 
at the position of a gas cell is also given. We use the criterion 
$\rho\ge 15\rho_c$ to distinguish gas elements belonging to haloes.
Since the dark matter is overabundant in haloes and its density
dominates anyway, this criterion is close to the one used by \cite{hai16}.
We neglect gas cells belonging to haloes when studying gas distribution
in the IGM.}
We project the electron distribution 
onto three mutually perpendicular walls of the simulation cube, using
top hat filter for each cell.
We choose 2D maps of $16~$k$\times16~$k resolution. The pixel size 
($\sim 5/h~\mathrm{kpc}$) corresponds to the comoving size of the densest
gas cells at $z=0$. ({\rm All cells belonging to the IGM have much
larger sizes.}) The resulting maps of projected free electron
distribution give probability distributions of 2D electron density
fluctuations $f_z(\delta_2)$ based on $3\times 16384^2$ "measurements" 
for  chosen redshifts.
By definition $\left<\delta_2\right>=0$. We find the range of fluctuations on
all considered maps to be $-0.86\le\delta_2\le$ {\rm 20.}. We use
histograms with logarithmic bins of width 0.01 dex for storing
and plotting probability distributions as functions of 
$-1\le\lg(1+\delta_2)\le$ {\rm 1.5}.} 

\begin{figure}
\includegraphics[width=\columnwidth]{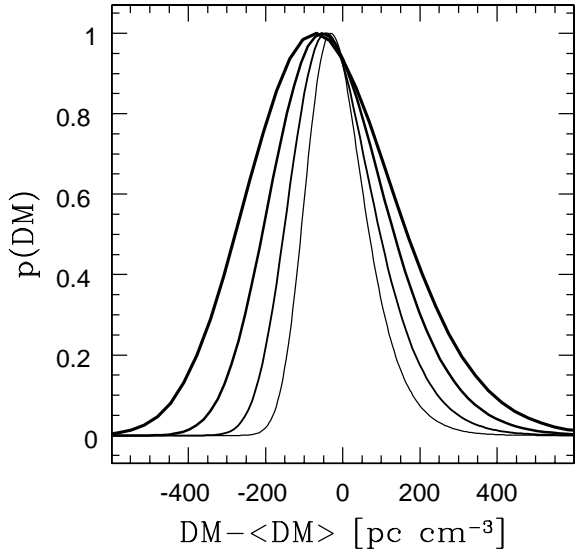}
\caption{
Probability distributions for the dispersion measures to sources at
the redshifts $z\approx 0.5$ (thin line, narrow plot)  
1, 2, and 4 (thick, wide). Values of $\left<DM\right>(z)$ are given in  
Table~\ref{cumulDM}.
}
\label{chosenDM}
\end{figure}

Using Illustris data we find {\rm the values of $f_{IGM}(z)$,
$f_{ion}(z)$, and $f_e(z)$ for a set of redshifts  
(compare the upper part of Table~\ref{fion}). Almost all the baryons belong
to the IGM at $z=5$ (not shown in the Table) and then their fraction
decreases monotonically to $f_{IGM}\approx 0.84$ at $z\approx 1$, 
to increase slightly at $z=0$ due to the gas ejection from haloes. 
Since $z=3$ IGM gas is almost
completely ionized with $f_{ion}\ge 0.99 $, so $f_e$ is very close to
$f_{IGM}$. At $z=5$ $f_{ion}\approx 0.93$ due to incomplete helium
ionization. }
The electron fraction parameter 
changes only by few per cent in the range $0\le z \le 5$, and we linearly
interpolate its value between $z=0$ and $z=1$, 1 and 2, etc where 
we know $f_e(z)$ based on Illustris data.} At $z>5$ electron concentration 
given by Illustris becomes nonphysically low and we do not exploit this 
range. On the other hand the results of
Illustris for $2<z<4$ are in good agreement with the observations of
Ly-$\alpha$ forest \citep{vog14a} and there the data can be trusted.
The plots of $d\left<DM\right>/dz$ in Fig.~\ref{avgDM} show
the changes in Illustris electron density (solid line) as compared to
the model assuming full ionization of all baryonic matter (dotted). 


{\rm We approximate the probability distributions at any $z$
interpolating between the distributions given by Illustris. For any 
$z$: $z_1 \le z \le z_2$ we assume
\begin{equation}
f_z(\delta_2)=\frac{z_2-z}{z_2-z_1}f_{z_1}(\delta_2)
             +\frac{z-z_1}{z_2-z_1}f_{z_2}(\delta_2)
\label{interpolation}
\end{equation}
where $z_1\in\{0,1,2,3,4\}$ and $z_2=z_1+1$. We check our assumption
comparing $f_{0.5}(\delta_2)$ based on Illustris data with interpolated 
$f_{01}(\delta_2)=0.5~f_0(\delta_2)+0.5~f_1(\delta_2)$. We show both
distributions in Fig.~\ref{interpolated}. We also perform
Kolmogorov-Smirnov test comparing the cumulative probability
distributions:
\begin{eqnarray}
F_z(\delta_2)&=&\int_{-\infty}^{\delta_2}~f_z(x)dx\\
\max_{\delta_2}(|F_{01}(\delta_2)-F_{0.5}(\delta_2)|)&=&0.0097
~~~\mathrm{at}~~~\delta_2=-0.45
\end{eqnarray}
The maximal difference between cumulative probability functions is much
lower than the critical value of K-S test ($\sim 1/\sqrt{N}$) for
$N=250$ bins.}

We have performed $10^8$ trial simulation calculating dispersion measure
for sources at the redshifts $0<z_S<5$, which 
in our approach corresponds to crossing 
up to 75 ($z_S\approx 5$) simulation cubes.
The dispersion measure  is calculated using Eq.~\ref{defDM3}
with $\delta_2$  at each cube chosen at random using
probability distributions {\rm like these shown in} Fig.~\ref{interpolated}.

The relative probability distributions of the simulated $DM$ to sources at the
chosen redshifts is shown in {\rm Fig.~\ref{chosenDM}}.  
In Table~\ref{cumulDM} we give the expected $DM$ value
expressed in units of $\mathrm{pc}~\mathrm{cm}^{-3}$ and its
standard deviation $\sigma_{DM}$ for chosen source redshifts. 
{\rm The values of higher moments of the probability distributions 
$\gamma_n\equiv\left<(DM-\left<DM\right>)^n\right>/\sigma_{DM}^n$ 
show that they are {\rm not} far from being Gaussian (in which case
$\gamma_3=0$ and $\gamma_4=3$).}


\section{Simulating FRBs observations and fitting cosmological parameters}.

We simulate observations of FRBs with known redshifts. 
The source redshifts are drawn at random from the distribution 
$f(z)\propto z\exp(-z)$ (\citealt{zhou14},\citealt{wal18})
limited  to the range $0.5\le z \le 4.5$. {\rm We also consider smaller
redshift ranges $0.5\le z \le 2.5$ and $0.5\le z \le 1.5$.}
(The redshift distribution $f(z)$ describes gamma ray bursts (GRBs).  
We do not assume any physical relation between GRBs and FRBs. The 
difficulty in measuring redshifts for both kinds of phenomena may be
similar, and FRBs may be related to stellar evolution as are GRBs.
We use the distribution of observed redshifts of a known class 
of sources only as an example.)

Following {\rm \cite{dz14}, \cite{gao14}} we assume, that the 
observations give the
dispersion measure which is a sum of contributions from inter-galaxy
medium, host galaxy,  and our Galaxy:
\begin{equation}
DM_{obs}=DM_{IGM}+DM_{host}+DM_{Galaxy}
\end{equation}
The contribution of our Galaxy to
the dispersion measure can be estimated based on local measurements 
and subtracted from the observed value.
{\rm As shown by simulations of \cite{yz16} the averaged value of the 
host galaxy
contribution can be estimated based on observations of close FRBs.
This result is based on the assumption that $\left<DM_{host}\right>$ 
does not significantly evolve which implies its $\propto 1/(1+z)$
contribution to $DM_{obs}$, while for the IGM part we have roughly 
$\left<DM_{IGM}\right>\propto z$ (compare Fig.~\ref{avgDM},
Table~\ref{cumulDM}).  Different redshift dependencies 
allow fits of both parts for a large enough sample of close bursts.
In the following we assume, that the average host contribution can be
subtracted from the observed $DM$, giving the approximate value of the
IGM part:}
\begin{equation}
DM_{IGM,obs}=DM_{obs}-\left<DM_{host}\right>/(1+z)-DM_{Galaxy}
\end{equation}
{\rm According to \cite{ten17}, who analyze the repeating FRB 121102 
and \cite{yang17}, who statistically investigate 21 FRBs, 
the value of $DM_{host}$ is high 
($\ge 200~\mathrm{pc}~\mathrm{cm}^{-3}$) and has large scatter. }
Following \cite{yz16}, \cite{wal18} we assume $DM_{host}$ to have normal
distribution. We adopt the value
$\sigma_{host}=50~\mathrm{pc}~\mathrm{cm}^{-3}$ for its standard
deviation. Thus our
simulated $DM(z_k)$, {\rm which represents its IGM part after
subtracting the Galaxy and averaged host contributions,}  
includes also an extra term which is normally
distributed:
\begin{eqnarray}
DM_k&=&DM(z_k)+\sigma_h(z_k)*\mathcal{N}(0,1)\\
\sigma_h(z_k)&=&\sigma_{host}/(1+z_k)
\label{simulDM}
\end{eqnarray}
where $DM(z_k)$ is given by Eq.~\ref{deltaDM} and $\mathcal{N}(0,1)$  
is the normal distribution with zero mean and unit variance.

We fit model parameters denoted $\Theta$ looking for the minimum of
\begin{equation}
\chi^2_{FRB}=\sum_k
~\frac{\left(DM_k-DM_0(z_k;\Theta)\right)^2}
{\sigma_k^2+\sigma_h^2(z_k)}
\label{chi2}
\end{equation}
which is a standard approach. $DM_0(z_k;\Theta)$ is a model prediction
of the dispersion measure to the source at the redshift $z_k$. 
{\rm Despite the fact that}  the distribution of simulated $DM_k$ 
is {\rm only weakly} non-Gaussian, {\rm we always investigate the
distribution of calculated $\chi^2$} and set its
critical value $\chi^2_{crit}$ such that in 95\% of cases
$\chi^2\le\chi^2_{crit}$. We accept fits with $\chi^2$ values below
critical. 

\begin{table}
  \caption{Electron fraction parameters}
  \begin{tabular}{ccccccc}
  \hline
&&& Illustris &&&\\
\hline
$z$   &   & 0 & 1 & 2 & 3 & 4 \\
$f_{IGM}$ & & 0.847 & 0.841 & 0.876 & 0.928 & 0.964\\ 
$f_{ion}$ & & 0.993 & 0.992 & 0.992 & 0.990 & 0.979\\
$f_e$     & & 0.841 & 0.835 & 0.869 & 0.918 & 0.944\\
\hline
 &&& Fits to $f_e$ &&&\\
\hline
100&4.5&.83$\pm$.10 & .83$\pm$.09 & .87$\pm$.11 & .92$\pm$.17 & .99$\pm$.32\\
400&4.5&.83$\pm$.05 & .83$\pm$.04 & .86$\pm$.06 & .90$\pm$.08 & .98$\pm$.14\\
100&2.5&.83$\pm$.07 & .83$\pm$.06 & .87$\pm$.08 &      &      \\
100&1.5&.83$\pm$.07 & .83$\pm$.07 & .87$\pm$.50 &      &      \\
$\sigma_{mod}$& & .008 & .010 & .011 & .008 & .012 \\
 \hline
\end{tabular}
\label{fion}
\end{table}

Illustris gives the concentration of electrons in space and its time
dependence for a single realization of simulations performed in a
cosmological model with given set of parameters. We do not have similar
information based on simulations for other universe models. The scenarios
of gravitational instability and reionization influence the 
distribution of electrons in a complicated way specific to each model. 
Results of Illustris show, that the {\rm fraction of baryons in the IGM
changes with time and} ionization is incomplete {\rm at early epochs so} 
working under the simplifying assumptions {\rm of $f_e\approx 1$ one} would
systematically overestimate model values of $DM$. Thus introduction of
at least one parameter describing the free electron fraction seems necessary. 
The FRBs tests to find $\Omega_B f_e$ has been proposed by
\cite{dz14}. \cite{sh18} propose calibration of the fraction of baryons
in the diffuse IGM. We follow these ideas checking the possibility of
estimating the free electron fraction in the IGM and its dependence on
the redshift. 

{\rm First we check whether it is possible to recover the values of
$f_e(z_i)$ ($z_i=0$, 1, 2, 3, 4) used in simulations and given in the
upper part of Table~\ref{fion} assuming other
model parameters ($H_0$, $\Omega_M$, $\Omega_\Lambda$, $\Omega_B$) to
have the values used in Illustris. (When dealing with
real data one would use their current {\it precision cosmology} values.)
Since $DM_0(z_k;\Theta)$ can be expressed as a linear combination of 
$f_e(z_i)$ the minimization of $\chi^2_{FRB}$ becomes a {\it linear 
least squares}  problem solvable with standard methods of the linear 
algebra.  Using our methods we
generate many samples of simulated FRBs drawing their redshifts from the
assumed distribution and use the range $0.5\le z_k\le 4.5$ with $1\le k \le N$.
For $N=100$ we encounter numerical problems ("singular matrix") in less
than one case per million. For any simulated sample the fitted parameters
may be unphysical ($f_e<0$) which is rare and in practice
happens only for $f_e(z=4)$ or have excessive values $f_e>1$. The
latter occurs when sources belonging to a sample happen to lie behind 
over-dense regions of space. On average the fitted $f_e$ values
reproduce their simulation counterparts. Their standard deviations
depend on the sample size $N$ in usual way $\propto 1/\sqrt{N}$ - compare
Table~\ref{fion}, but are rather high. Averaged electron density for
$0\le z\le 1$ depends on $f_e(z=0)$ and $f_e(z=1)$ in our
approach and these two parameters influence the expected dispersion
measure for all sources in a sample, while $f_e(z=4)$ has an impact
only on sources with $z_k\ge 3$, which are less numerous ($0.22~N$ on
average). Thus the estimate of $f_e$ at higher redshifts has a
larger standard deviation. }

{\rm Using samples with smaller redshift range ($0.5\le z \le 2.5$, 
$0.5\le z \le 1.5$ ) we try to fit the electron fraction 
parameters for $z=0$, 1, and 2. In the first case the fits are as good 
as for samples with larger redshift ranges, but in the latter case 
$f_e(z=2)$ is practically unconstrained despite the fact that the expected 
values of $DM$ for all sources at $z>1$ do depend on this parameter.
Examining Table~\ref{fion} one can see that to fit $f_e$ with the 
accuracy of one per cent samples with $N\sim 10^4$ are needed. }

{\rm In the analysis above we have neglected the uncertainty resulting
from using wrong values of cosmological model parameters when 
fitting $f_e(z_i)$.  We assume four parameters $p_i$
($p_i\in\{H_0, \Omega_B, \Omega_M, \Omega_\Lambda\}$) to have values known
from other studies with errors $\sigma_i$. Neglecting correlations
between $p_i$ we would get:
\begin{equation}
\sigma^2_{model}=\sum_{i=1}^4~\left(\frac{\partial f_e}
{\partial p_i}\right)^2~\sigma_i^2
\label{uncert}
\end{equation}
The recent estimates of cosmological parameters are given in Planck
Results 2018 (\citealt{pl18}). According to this study $H_0$ is weakly
anti-correlated with $\Omega_B$. On the other hand the measured quantity 
$DM \propto f_e*\Omega_B*H_0$ (compare Eqs.~\ref{nel0},~\ref{defDM3}),
so Eq.~\ref{uncert} overestimates uncertainty resulting from errors in
these two parameters. The universe model is practically flat (ibid.) so
$\Omega_M$ and $\Omega_\Lambda$ are strictly anti-correlated. We examine
the influence of their values on electron fractions by repeating
calculations for $\Omega_M=\Omega_{MI}+\Delta$ and
$\Omega_\Lambda=\Omega_{\Lambda I}-\Delta$, where
$\Delta=\sigma_{\Omega_M}=\sigma_{\Omega_\Lambda}$. We add the estimated
uncertainty in quadrature to uncertainties from the other two parameters
getting values shown in the last row of Table~\ref{fion}. 
Because we have neglected the
weak anti-correlation between $H_0$ and $\Omega_B$ the numbers in the
table should be treated as upper limits. This shows that the uncertainty
of fitted electron fractions resulting from the uncertainty of the used
{\it precision cosmology} model is of the order of one per cent.}

\begin{figure}
\includegraphics[width=\columnwidth]{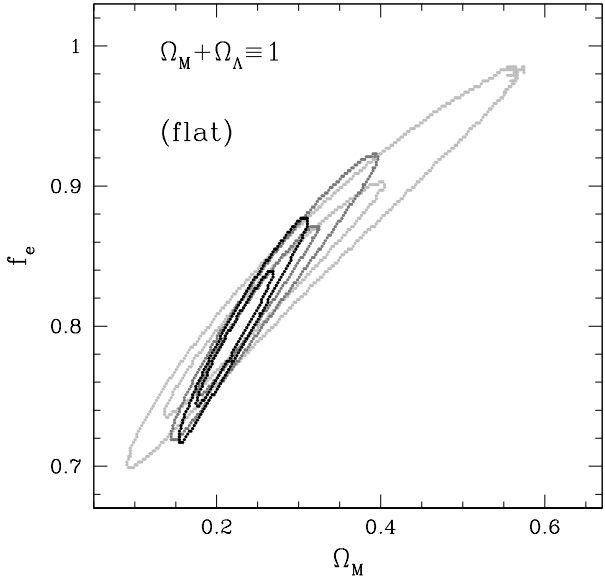}
\caption{
Distribution of fits to a flat $\Lambda$CDM model. Three redshift ranges
($0.5\le z\le 4.5$ - black; 
 $0.5\le z\le 2.5$ - dark gray;
 $0.5\le z\le 1.5$ - light gray) of the FRBs have been
considered. The 68\% and 95\%  confidence regions are shown, 
based on $\sim 4\times 10^6$ simulated FRBs samples for each redshift range.
}
\label{flation}
\end{figure}

{\rm Next we check the possibility of testing both the electron fraction and
the geometry of the universe model using FRBs. 
Since the dispersion measure is directly proportional to the product of
$f_e$, $H_0$, and $\Omega_B$ (Eqs.~\ref{nel0},\ref{defDM3}) 
placing any limits on 
these parameters separately is impossible. We simplify our approach
looking only for the expected electron fraction assuming $H_0$ 
and $\Omega_B$ to be known independently. The dispersion measure
to  source at a given redshift $z_S$ 
depends on the electron fraction in the redshift
interval $0\le z \le z_S$ in proportion to }
\begin{equation}
\left<f_e\right>(z_S)=\frac{\int_0^{z_S}f_e(z)\frac{(1+z)dz}{h(z)}}
            {\int_0^{z_S}\frac{(1+z)dz}{h(z)}}
\label{avgfe}
\end{equation}
{\rm The averaged electron fraction $\left<f_e\right>$ is slowly and 
monotonically changing from 0.825 at $z=0.5$ to 0.845 at $z=4.5$.
Averaging over the source redshift distribution
\begin{equation}
\left<\left< f_e\right>\right> 
=\frac{\int_{z_{min}}^{z_{max}}\left<f_e\right>(z_S)f(z_S)dz_s}
      {\int_{z_{min}}^{z_{max}}f(z_S)dz_s}
\label{avgavgfe}
\end{equation}
we get the expected $\left<\left<f_e\right>\right>$ values which are
almost the same for the three ranges of sources redshifts considered
(see Table~\ref{flatelel}). }

{\rm Despite simplifications we have been unable to place any interesting
constraints on $\Lambda$CDM model with three free parameters
($\Omega_M$, $\Omega_\Lambda$, $f_e$). The fits give large and irregular
confidence regions with a substantial fraction of solutions grouped
near the boundaries of the considered parameters region. (We limit the
possible values of the parameters to
$\Omega_{BI}\le\Omega_M\le\Omega_{BI}+1$, $0.3\le\Omega_\Lambda\le 1.3$,
$0.3\le f_e\le 1.3$.)}

\begin{figure*}
\includegraphics[width=176mm]{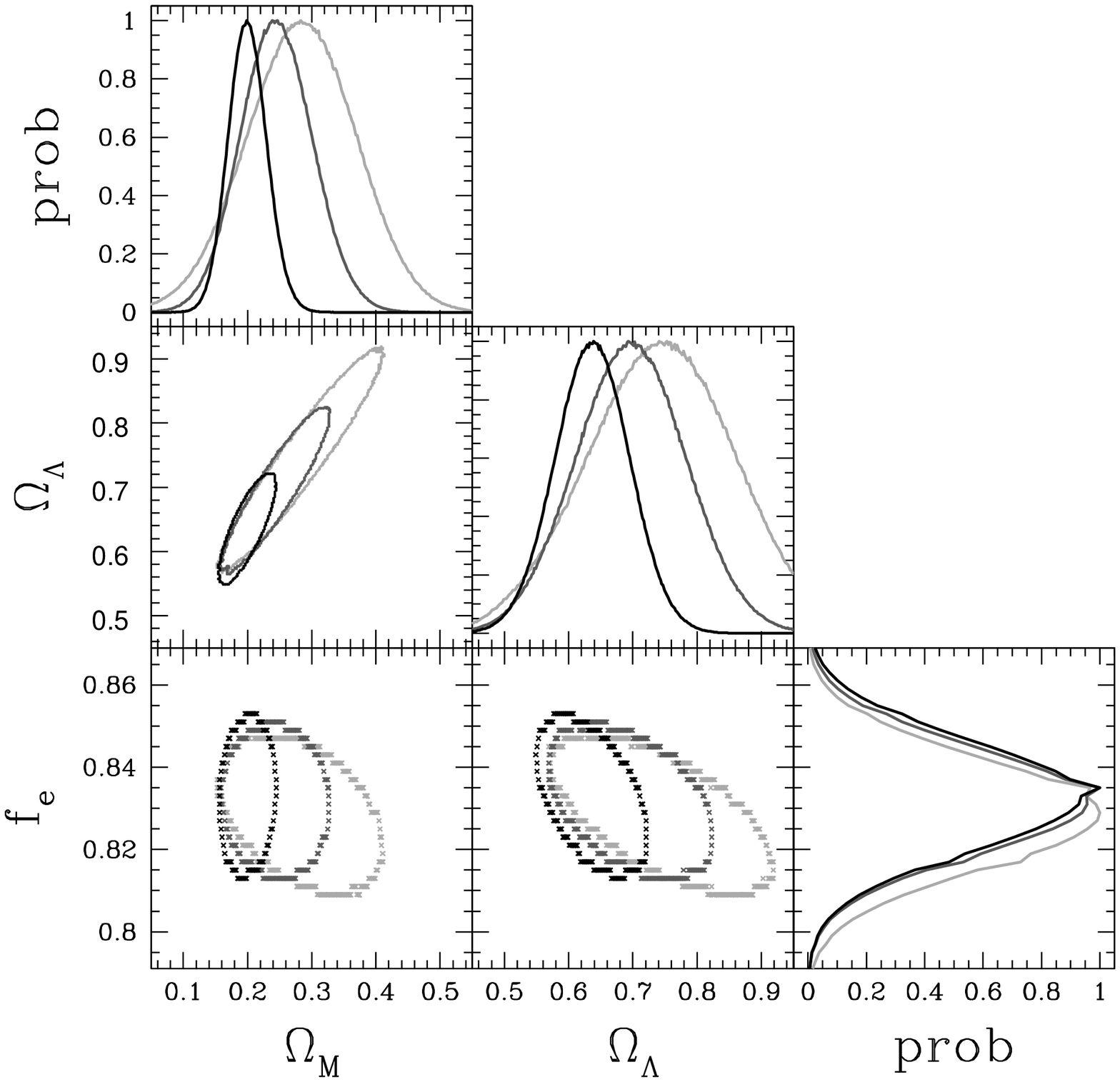}
\caption{Confidence regions for $\Lambda$CDM model parameters and
averaged electron fraction based on combined data (synthetic SN$^\prime$ Ia
samples and our simulated FRBs samples). We plot boundaries of
regions of 68\% confidence level. Results for the samples with the
largest redshift range ($0.5\le z_S\le 4.5$) are plotted with black lines,
for $0.5\le z_S\le 2.5$ - dark gray, and for $0.5\le z_S\le 1.5$ - light
gray.   
}
\label{all}
\end{figure*}

\begin{table}
  \caption{Fits to flat $\Lambda$CDM models}
  \begin{tabular}{cccccc}
  \hline
   N &$z_{max}$& $\Omega_M$ & $f_e$ & $f_e^\prime$ & $\left<\left<f_e\right>\right>$ \\
\hline
100&4.5& 0.23$\pm$0.03 &0.80$\pm$0.03 & 0.842$\pm$0.008 & 0.847 \\
100&2.5& 0.26$\pm$0.05 &0.82$\pm$0.04 & 0.833$\pm$0.009 & 0.841 \\
100&1.5& 0.29$\pm$0.10 &0.83$\pm$0.06 & 0.829$\pm$0.010 & 0.838 \\
 \hline
\end{tabular}
\label{flatelel}
\end{table}

We show the results for flat $\Lambda$CDM models in Fig.~\ref{flation}
and in Table~\ref{flatelel}. 
We have considered three ranges of FRBs redshifts in our simulations 
($0.5\le z_S\le 4.5$, $0.5\le z_S\le 2.5$, and $0.5\le z_S\le 1.5$).
In each case we have simulated $4\times 10^6$ "observed" FRBs samples.
For each sample we have fitted a flat $\Lambda$CDM cosmological model.
We have obtained histograms of fitted parameter values
$N(\Omega_M,f_e)$ for various ranges of source redshifts.
We find, that even a wide redshift range of FRBs ($0.5\le z_S\le 4.5$) 
gives only a rough estimate of parameters. The fitted $\Omega_M$ and $f_e$
are strongly correlated with $\rho\approx 0.97$. The averaged result of
many fits depends on the sample's redshift range and is biased.

{\rm Fixing both density parameters ($\Omega_M=\Omega_{MI}$, 
$\Omega_\Lambda=\Omega_{\Lambda I}$) we get estimates of sample averaged
electron fraction, which we denote $f_e^\prime$ in Table~\ref{flatelel}.
Their values agree with the analytical prediction
$\left<\left<f_e\right>\right>$. This shows that in a fixed cosmological
model one can get the redshift and sample averaged electron fraction
with the uncertainty of $\sim$ two per cent using a sample of one hundred
FRBs (plus one per cent uncertainty of the model). }


\section{Tests based on FRBs and other data}

{\rm Limited possibilities of FRBs tests suggest to use also some other data
to make useful constrains on the models, obtaining at least cosmological 
density parameters and electron fraction simultaneously. }

As an example we combine simulated FRBs and SNe Ia samples. 
We use the SCP ``Union'' SN Ia data
(\citealt{kow08}) {\rm as a basis of obtaining our synthetic samples.}
The ``Union'' sample contains 307 {\it usable} lightcurves. We have 
downloaded the data in the form  $\{z_j,\mu_j,\sigma_j\}$ redshift -- 
distance modulus -- its estimated error. {\rm The best fit of a $\Lambda$CDM
model to the real SNe data gives cosmological parameters which are
within 1-sigma from Illustris parameters but are different. To check
whether our tests reproduce cosmological parameters used in simulations
we replace the ``Union'' SN Ia data by synthetic samples SN$^\prime$ 
with the same redshifts and estimated errors but {\it corrected} 
distance moduli: }
$\mu_j^\prime$ with Gaussian noise:
\begin{equation}
\mu_j^\prime=
5\lg\left(\frac{d_L(z_j;\Omega_{MI},\Omega_{\Lambda I})}{10~\mathrm{pc}}\right)
+\sigma_j\mathcal{N}(0,1)
\label{modulus}
\end{equation}
where $d_L$ is the luminosity distance and Illustris cosmology
parameters are used.

We consider  $\Lambda$CDM models with {\rm two} free 
parameters ($\Omega_M$ and $\Omega_\Lambda$) while the Hubble constant
and baryon density are fixed at Illustris values. {\rm The averaged
electron fraction $f_e$, which does not influence the geometry of the
universe model or the SN$^\prime$ data fits, is necessary to model FRBs.}
The SN$^\prime$ part of the $\chi^2$ reads: 
\begin{eqnarray}
\chi^2_{SN}&=&\sum_{j=1}^{307}~
\frac{(\mu_j^\prime-\mu_0(z_j;\Omega_M,\Omega_\Lambda))^2}{\sigma_j^2}\\
\mu_0(z_j;\Omega_M,\Omega_\Lambda)&=&
5\lg\left(\frac{d_L(z_j;\Omega_M,\Omega_\Lambda)}{10~\mathrm{pc}}\right)
\end{eqnarray}
where  $\mu_0$ is a model predicted distance modulus. 

Now we look for cosmological parameters which minimize $\chi^2$ for a
combined FRB plus SN$^\prime$ data. For each FRB and SN$^\prime$ 
data simulation we find the minimum of:
\begin{equation}
\chi^2=\chi^2_{FRB}(\Omega_M,\Omega_\Lambda,f_e)+
       \chi^2_{SN}(\Omega_M,\Omega_\Lambda)
\end{equation}
Since there are 307 SN Ia in the SN Union sample, we use simulated FRBs
data samples consisting of 300 sources each. We examine many combined
sets.  In Fig.~\ref{all} we show confidence regions for
$\Omega_M$, $\Omega_\Lambda$, {\rm and sample averaged $f_e$ projected into 
two or one dimension. Table~\ref{combined} gives the fitted parameter
values with 1D 68\% confidence regions.} 

\begin{table}
  \caption{Fits of $\Lambda$CDM and electron fraction models to combined
   synthetic samples of SN$^\prime$e and FRBs}
  \begin{tabular}{lcccc}
  \hline
  model & $z_{max}$ & $\Omega_M$ & $\Omega_\Lambda$ & $f_e$ \\
\hline
 Illustris     && 0.2726          & 0.7274          &  \\
 SN         && 0.345$\pm$0.099 & 0.863$\pm$0.245 & \\
 SN$^\prime$ \& FRBs &4.5& 0.20$\pm$0.03 & 0.64$\pm$0.06 & 0.832$\pm$0.013\\
 SN$^\prime$ \& FRBs &2.5& 0.24$\pm$0.06 & 0.70$\pm$0.09 & 0.831$\pm$0.013\\
 SN$^\prime$ \& FRBs &1.5& 0.28$\pm$0.09 & 0.74$\pm$0.12 & 0.828$\pm$0.014\\
 SN$^\prime$ \& FRBs flat &4.5 & 0.254$\pm$0.013 & 1-$\Omega_M$& 0.825$\pm$0.013\\
  \hline
\end{tabular}
\label{combined}
\end{table}

{\rm The tests based on combined data allow to constrain both cosmological
density parameters ($\Omega_M$, $\Omega_\Lambda$) but their accuracy 
is still far from the present {\it precision cosmology} standards.
On the other hand our simulation shows that using few hundred of FRBs 
with another moderate precision test one is able to measure the averaged
electron fraction with uncertainty of $\sim$ two per cent.}


\section{Discussion}

{\rm We have analyzed the evolution of the distribution of ionized gas 
in {\rm the IGM} using the results of the Illustris simulation 
\citep{vog14a, vog14b}. According to {\rm our calculations the fraction
of gas in the IGM changes from $f_{IGM}\approx 1$ at $z=5$ to 0.85 at
$z=0$. This result is different from \cite{hai16}, who obtain
$f_{IGM}\approx 0.77$ at $z=0$. For $z<3$ the IGM gas is almost
completely ionized, and free electron fraction $f_e\approx f_{IGM}$. } 
The estimate of \cite{sh18} who claim that the diffuse baryons
constitute 0.6$\pm$0.1 of all gas is {\rm based on different reasoning.}

{\rm The spatial distribution of ionized gas and its evolution given by 
Illustris allows us to simulate calculations of  the dispersion measure 
along any LOS and obtain its standard deviation as a function of the source
redshift. 
Our results on averaged $DM$ to a source at given redshift
(Fig.~\ref{avgDM}, Table~\ref{cumulDM}) are in {\rm rough} agreement 
with other authors (e.g. \citealt{zhou14}, \citealt{dz14}), 
{\rm and with the approximate formula of \cite{zhang18}:}
\begin{equation}
DM_{IGM} \approx 855~\mathrm{pc}~\mathrm{cm}^{-3}~z
\end{equation} 
At $z=1$ {\rm we get} $\sigma_{DM}$ {\rm which is about two times lower
as compared to the results} 
of \cite{mcq14} based on simulations. {\rm Only if all gas from
Illustris except} {\it stellar cells} {\rm is included in calculations
we get the agreement with his $\sigma_{DM}(z=1)$ value.} 
The shape of our $\sigma_{DM}(z)$ 
dependence is different and resembles his analytical results.

{\rm The dispersion measure to a given source is directly proportional to the
density of free electrons along LOS. In our approach we model it using
the electron fraction parameter $f_e(z)$ (Eq.~\ref{deffe}). In
simulations we use its values at $z=0, 1, ..., 5$ and linearly
interpolate between them. Following \cite{dz14}, \cite{sh18} and others
we check the possibility of finding the history of ionization
using a sample of observed FRBs and assuming cosmological parameters 
($H_0$, $\Omega_B$, $\Omega_M$, $\Omega_\Lambda$) to be known exactly.
We show, that to get the the dependence
of the electron fraction on the redshift with one per cent accuracy,
samples of $N\sim 10^4$ events are needed (compare Table~\ref{fion}). 
The present errors in cosmological parameters values (e.g. \citealt{pl18})
introduce another one per cent uncertainty. On the other hand the
redshift and sample averaged electron fraction can be constrained
with accuracy better than two per cent based on $N=100$ FRBs in a fixed
cosmological model (compare Table~\ref{flatelel}). The dependence on
cosmological parameter errors is the same as above. }

{\rm We have shown, that using a sample of three hundred FRBs and a sample
resembling SN Ia Union Sample (\citealt{kow08}) we are able to
constrain the averaged electron fraction parameter with the accuracy
of $\sim$ two per cent (compare Table~\ref{combined}). }

The cosmological tests based on large future samples of FRBs are possible,
but are unlikely to give the accuracy of the present day precision cosmology. 
This conclusion with more details has already been reached by \cite{wal18}.
There are degeneracies between derived universe model parameters, which 
can be removed only by using other type of data, as shown by our 
consideration of joint FRBs and SN$^\prime$ Ia tests. 
The observations of FRBs with measured redshifts may give valuable data 
on the distribution of free electrons in space also on cosmological
scales (compare \citealt{sh18}). {\rm Since the dispersion measure
depends on the electron distribution between the source and the
observer, the dependence of the electron fraction on the redshift can, 
at least in principle, be investigated. 
In this aspect samples of FRBs with known redshifts may be
a better target than Cosmic Microwave Background observations yielding
Thompson optical depth $\tau_e$ and Suniajev-Zeldovich $y$ parameter
characterizing the electron population between the observer
and the Last Scattering Surface (see e.g. \citealt{pl18}). }

\section*{Acknowledgments}

We thank the Anonymous Referee for his constructive comments, which helped to
improve the paper.
The Illustris Simulation databases used in this paper
and the web application providing on-line access to them were constructed
as part of the activities of the German Astrophysical Virtual Observatory.

\end{document}